\documentclass[aps,prl,twocolumn,superscriptaddress,showpacs,floatfix,nobibnotes]{revtex4}
\usepackage{epsfig}
\usepackage{epstopdf}

\usepackage{graphicx}
\usepackage{longtable}
\usepackage{CJK}
\usepackage{color}

\usepackage{mathptmx, courier, pifont}
\usepackage[scaled=0.92]{helvet}
\usepackage[T1]{fontenc}
\usepackage{textcomp}

\usepackage{fancyhdr}

\begin{document}

\begin{CJK*}{GB}{}

\title{Shape phase transition in the odd Sm nuclei: effective order parameter and odd-even effect}

\author{Yu Zhang}
\affiliation{Department of Physics, Liaoning Normal University,
Dalian 116029, P. R. China}

\author{Xin Guan}
\affiliation{Department of Physics, Liaoning Normal University,
Dalian 116029, P. R. China}

\author{Yin Wang}
\affiliation{Department of Physics, Chifeng University, Chifeng
024000, P. R. China}

\author{Yan Zuo}
\affiliation{Department of Physics, Liaoning Normal University,
Dalian 116029, P. R. China}

\author{Li-na Bao}
\affiliation{Department of Physics, Liaoning Normal University,
Dalian 116029, P. R. China}

\author{Feng Pan}\affiliation{Department of Physics, Liaoning Normal University,
Dalian 116029, P. R. China}

\date{\today}

\date{\today}

\begin{abstract}

Some binding-energy-related quantities serving as effective order
parameters have been used to analyze the shape phase transition in
the odd Sm nuclei. It is found that the signals of phase transition
in the odd Sm nuclei are greatly enhanced in contrast to the even Sm
nuclei. A further analysis shows that the transitional behaviors
related to pairing in the Sm nuclei can be well described by the
mean field plus pairing interaction model, with a monotonic decrease
in the pairing strength $G$.
\end{abstract}
\pacs{21.60.Ev, 21.10.Re, 64.70.Tg}

\maketitle

\thispagestyle{fancy} \fancyhead{} \lhead{Submitted to Chinese
Physics C} \chead{} \rhead{} \lfoot{} \cfoot{\thepage} \rfoot{}
\renewcommand{\headrulewidth}{0pt}
\renewcommand{\footrulewidth}{0.7pt}

\end{CJK*}

\begin{center}
\vskip.2cm\textbf{I. Introduction}
\end{center}\vskip.2cm

Quantum phase transitions in nuclei have attracted a lot of
 attention from both experimental and theoretical
 perspectives~\cite{Iachello2000,Iachello2001,Jolie2001,Jolie2002,Iachello2004,Sun2008,Mu2005,Zhangjinfu2003,
 Zhangdali2003,Zhangdali2002,Zhangyu,Dai2011,WC1983,Liu2006,Zhang2008,Liu2007,Zhang2013,Zhang2013II,Luo2006,Li2009,
 Pan2003,Casten2007,Cejnar2009,Cejnar2010,IachelloBook87},
 since they provide new insights into understanding the evolution of
 nuclear properties. The quantum phase transition is not of
 the usual thermodynamic type, but related to the equilibrium shape
 changes in the ground state of nuclei at zero temperature. It is thus also referred to as
 the shape phase transition or ground state phase transition, though the concept can also be applied to excited states.
 Evidences of the shape phase transition in nuclei are signaled
 experimentally through a sudden change in the properties of the
 ground state. An excellent example is provided by a set of even Sm isotopes, of which the evolution of its
 properties can be identified as the first-order shape phase
 phase transition experimentally~\cite{Casten2007}. On
 the other hand, odd-A nuclei can be approximately considered as
 systems with an even-even core coupled to a single valence nucleon. It
 is thus expected that the properties of odd-A nuclei should be definitely
 affected by the shape phase transition emerging along the related
 odd isotope or isotone chains.

 Currently, analyses of the phase transitions in
 nuclei are mostly focused on even-even systems and have been carried out in the frame
 of phenomenological geometrical models of nuclear potential~\cite{Iachello2000,Iachello2001}, or
 algebraic models of nuclear structure~\cite{IachelloBook87} since the phase transitions in the intermediate and heavy mass
 region, such as the mass number $A\sim150$ region,
 are often out of reach of the microscopic shell models.
 In addition, phase transitions in odd-A nuclei may be more
 difficult to describe due to much more complicate dynamical
 situations~\cite{Iachello2011} in contrast to the adjacent even-even species.
 However, if one only wants to emphasize some special aspects of nuclear phase
 transitions, the shell model (under some approximations) is still applicable to give a microscopic
 analysis of the phase transition in the $A\sim150$ region. The
 purpose of this work is to give a microscopic analysis of the
 shape phase transition in the odd Sm nuclei in terms of the
 effective order parameters and odd-even effects.

\begin{center}
\vskip.2cm\textbf{II. Effective order parameter}
\end{center}\vskip.2cm

\begin{figure}
\begin{center}
\includegraphics[scale=0.30]{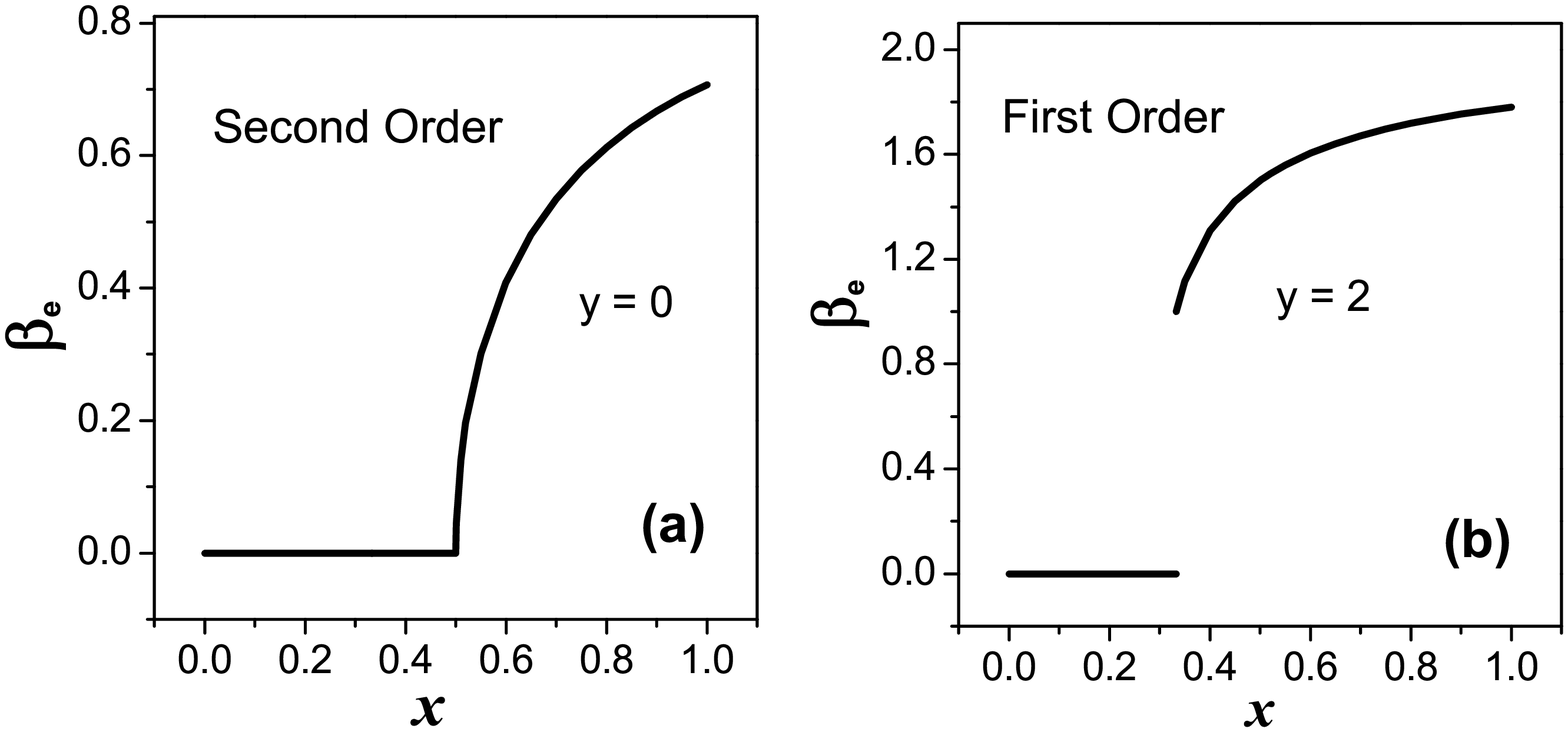}
\caption{(Color online) he order parameter $\beta_\mathrm{e}$ as a
function of $x$ for $y=0$ and $y=2$, respectively. \label{F01}}
\end{center}
\end{figure}

One way of addressing quantum phase transitions is to resort to the
 potential energy approach. To define phase transitions in theory, it
 is convenient to consider a schematic ``Landau''
 potential~\cite{Iachello2004} written as
 \begin{equation}\label{V}
 V(\beta)=\beta^2+x[(1-\beta^2)^2-y\beta^3],~~~\beta\geq0\, ,
 \end{equation}
 with two control parameters $0\geq x\geq1$ and $y\geq0$. This kind
 of potential may be formally derived from the interacting boson
 model~\cite{IachelloBook87}, which has been widely used to study
 quantum phase transitions in nuclei. It can be
 proven that the system has a second order phase transition at
 $x=x_c=1/2$ when $y=0$ because the minima of $V(\beta)$,
 $V_{\mathrm{min}}$, and $\frac{\partial V_{\mathrm{min}}}{\partial
 x}$ are continuous, but $\frac{\partial^2 V_{\mathrm{min}}}{\partial
 x^2}$ is discontinuous. More generally, the system will show a first
 order phase transition as a function of $x$ for any fixed value of
 $y>0$. For example, one can show that $V_{\mathrm{min}}$ is
 continuous but $\frac{\partial V_{\mathrm{min}}}{\partial x}$ is
 discontinuous at $x=x_c=1/3$ when $y=2$, which indicates the first
 order phase transition occurring at $x_c$. The potential~(\ref{V})
 can be considered as a simplified phenomenological nuclear potential
 surface varying as the function of the deformation $\beta$, which indicates that one
 could take $\beta_{\mathrm{equilib}}=\beta_\mathrm{e}$ to be the
 order parameter. As shown for the cases considered in
 Fig.~\ref{F01}, the order parameter $\beta_e$ changes continuously
 as a function of $x$, with the first derivative being discontinuous
 at $x_c=1/2$ if $y=0$, corresponding to the second-order phase
 transition, while $\beta_e$ jumps abruptly from $0$ to $1$ at
 $x_c=1/3$ when $y=2$, corresponding to the first-order phase
 transition. It seems that both $\beta_e$ and $V_{\mathrm{min}}$, which
 may correspond to the ground state deformation and energy respectively,
 can be used to identify the shape phase transition.
However, for real nuclei, things are considerably more complicated
 partly because $\beta_e$ is not an obervable. In fact, instead of
 $\beta_e$, the so called effective order
 parameters~\cite{Iachello2004} (observables
 sensitive to shape phase transitions occurring within nuclei) are often used
 to identify nuclear shape phase transitions and in some cases
 even determine their orders. The typical effective order parameters
 include the isomer shifts defined as $v=c[\langle
 r^2\rangle_{0_2}-\langle r^2\rangle_{0_1}]$ and
 $v^\prime=c^\prime[\langle r^2\rangle_{2_1}-\langle
 r^2\rangle_{0_1}]$~\cite{Iachello2004} with $c$ and $c^\prime$ being
 the scale parameters, the $B(E2)$ ratio $B(E2;(L+2)_1\rightarrow
 L_1)/B(E2;2_1\rightarrow0_1)$~\cite{Liu2007}, and the
 energy ratio $E_{L_1}/E_{0_2}$~\cite{Bonatsos2008}, {\it
 etc}. Most of the effective order parameters are related to the quantum
 numbers of excited states, which makes it particularly difficult use them to identify the phase transitions in odd nuclei. In contrast, the
 binding-energy-related quantities may serve as qualified effective
 order parameters to identify the phase transitions in both even-even
 and odd nuclei~\cite{Zhang2013} since their values only depend on the number of
 nucleons, and their experimental data are also relatively abundant.
 On the other hand, the number of nucleons in nuclei is
 finite and the phase transitional behavior will be
 muted due to the finiteness of the system~\cite{Iachello2004}. Specifically, instead of
 a discontinuity, sudden changes or flattening may be shown by the
 effective order parameters in nuclear shape phase transitions~\cite{Iachello2004,Casten2007}.

\begin{center}
\vskip.2cm\textbf{III. Two-neutron separation energy}
\end{center}\vskip.2cm

\begin{figure}
\begin{center}
\includegraphics[scale=0.30]{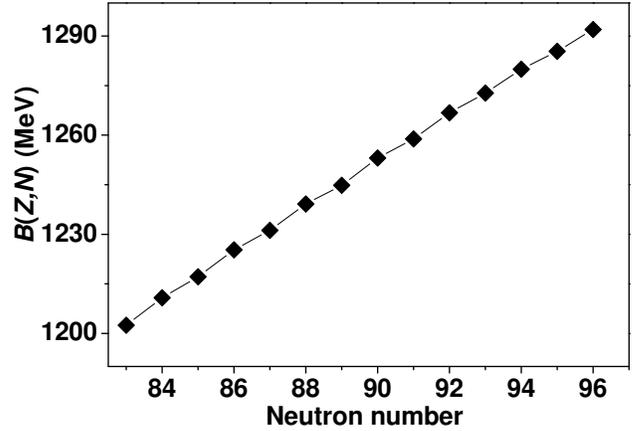}
\caption{Total binding energy $B(Z,N)$ for the Sm nuclei (taken from
\cite{Internet}) shown as functions of the neutron number.
\label{F0}}
\end{center}
\end{figure}

\begin{figure}
\begin{center}
\includegraphics[scale=0.35]{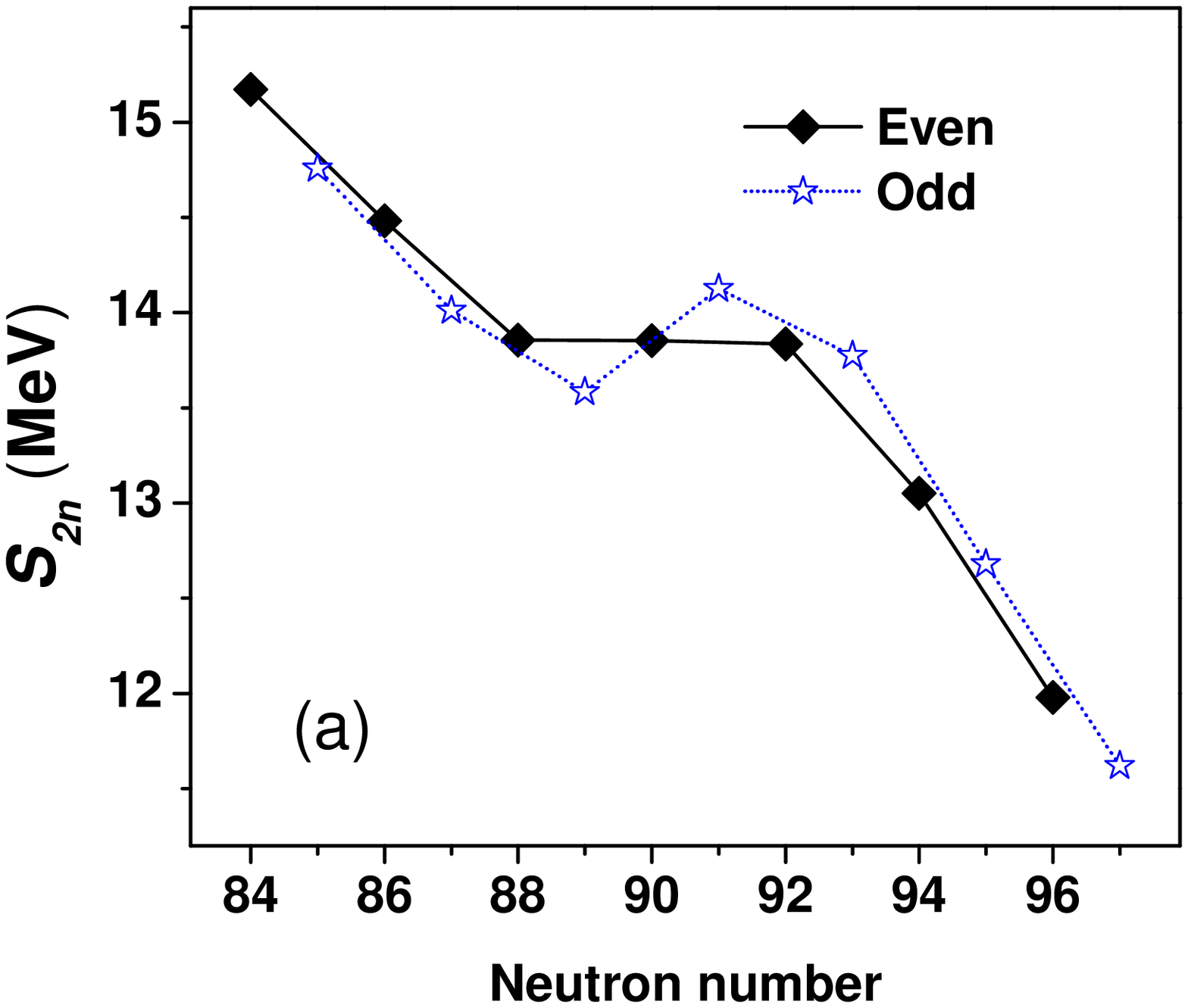}
\includegraphics[scale=0.35]{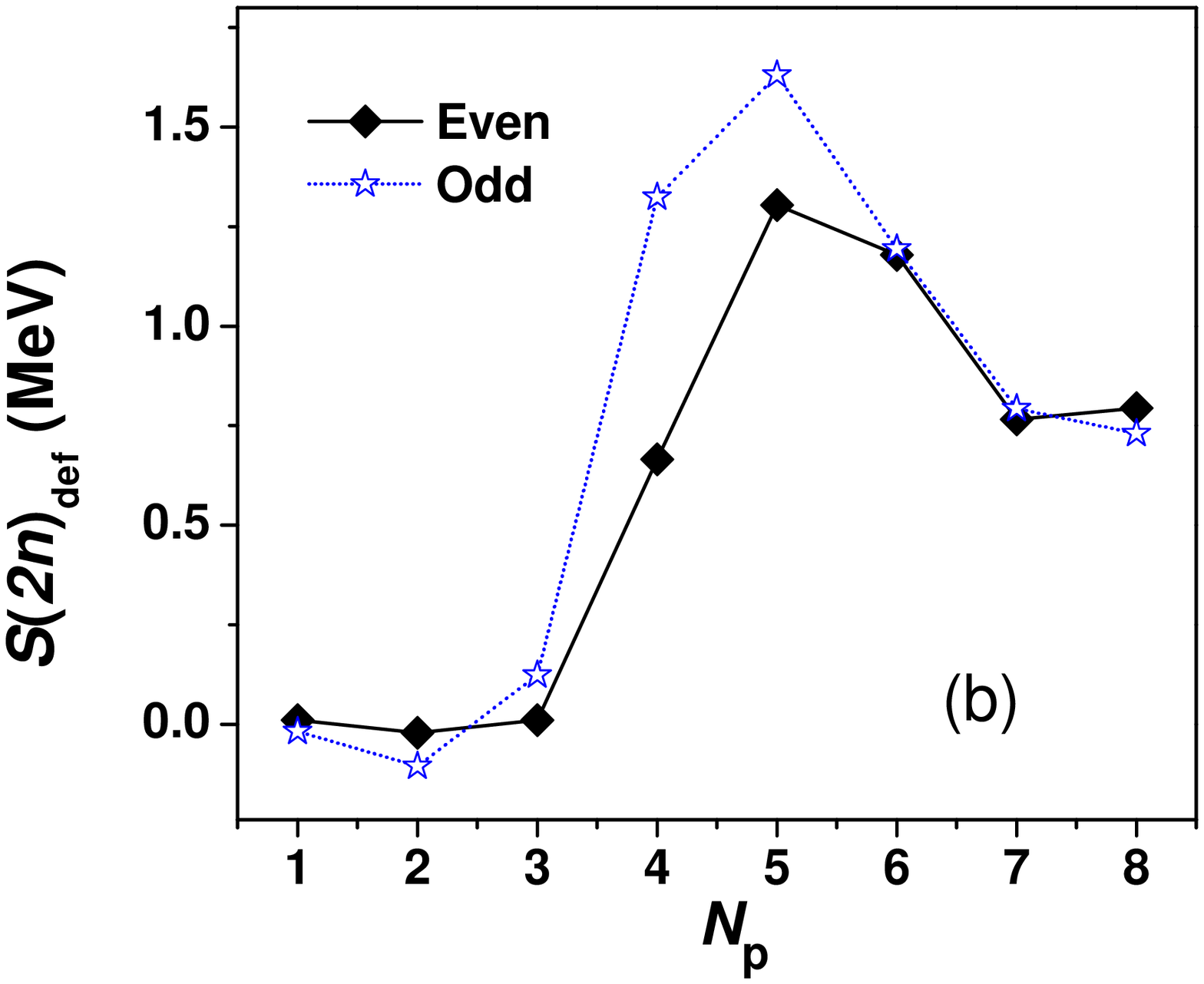}
\caption{The two-neutron separation energies, $S_{2n}$ and their
deformed part $S(2n)_{\mathrm{def}}$ for both the even and odd Sm
nuclei are shown along the isotopes chain. The experimental data are
taken from \cite{Internet}. \label{F1}}
\end{center}
\end{figure}

For an atomic nucleus, the most basic
 characteristic is nuclear mass or binding energy. The total binding energy $B(Z,N)$ for a nucleus
 with proton number $Z$ and neutron number $N$ is defined
 as~\cite{Ring1980}
 \begin{equation}
 B(Z,N)=ZM_\mathrm{p}+NM_\mathrm{n}-M(Z,N)\, ,
 \end{equation}
 where $M_\mathrm{p}$ is the proton mass, $M_\mathrm{n}$ is the neutron mass, and $M(Z,N)$ denotes the nuclear mass.
 In experiments, data about the binding energy $B(Z,N)$ is also
 abundant in contrast to other observables. One may expect that the shape phase transition in an isotope should be reflected by
 the evolution of the binding energy, which for the Sm isotopes is shown in Fig.~\ref{F0}.
 However, the varied energy scale of the total binding energy $B(Z,N)$ in an isotope chain is
 too large ($\sim10^2$ MeV) so that the signals of phase transition that are expected to
 appear around $N=90$ have been completely hidden
 behind the linear behavior of $B(Z,N)$ as shown in Fig.~\ref{F0}.
Thus one has to resort to other quantities related to the binding
energy in order to identify the shape phase
 transition in an isotope chain. For even-even nuclei, the two-neutron separation energy $S_{2n}$, which is defined
 as~\cite{Ring1980}
 \begin{equation}
 S_{2n}=B(Z,N)-B(Z,N-2)\, ,
 \end{equation}
 can be considered as a
 primary and direct signature of the emergence of
 the shape phase transition~\cite{Casten2007,Cejnar2010}.
 For odd-A nuclei, $S_{2n}$ can also serve as a qualified effective order parameter for identifying the ground phase transitions~\cite{Zhang2013}.
 For a set of isotopes, the two-neutron separation energy $S_{2n}$
 may be rewritten as a smooth contribution that is linear with the number
 of valence neutron pairs $N_p$, plus a contribution from the
 deformation~\cite{Iachello2011}
 \begin{equation}\label{S2n}
 S_{2n}=-A-B N_p+S(2n)_{\mathrm{def}}\, ,
 \end{equation}
 where $A$ and $B$ are the parameters.
 To emphasize the occurrence of the phase transition in the Sm
 nuclei, the experimental data of $S_{2n}$ for both the even and odd Sm isotopes~\cite{Zhang2013}
 are shown in Fig.~\ref{F1}, where the deformation contributions $S(2n)_{\mathrm{def}}$
 are also shown to reveal the odd particle (single valence nucleus) effects on the phase transition.
It is easy to know from Eq.~(\ref{S2n}) that the deformation
contributions $S(2n)_{\mathrm{def}}$ can be obtained from the data
by subtracting a term linear with the number of valence neutron
pairs $N_p$. Concretely, the results of $S(2n)_{\mathrm{def}}$ are
obtained from the data~\cite{Internet} fitted with $A=-19.8$ MeV and
$-19.4$ MeV for the even and odd Sm nuclei respectively, and
$B=0.66$ MeV according to (\ref{S2n}). As clearly seen from
Fig.~\ref{F1}(a), a noticeable feature is the sudden flattening near
the neutron number $N=90$ shown by $S_{2n}$ for the even Sm
isotopes. Based on the analysis given in \cite{Casten2007}, the
sudden flattening indicates the first order phase transition
emerging in the corresponding isotopes. It is even more interesting
to find that a similar or even more pronounced change appears in
$S_{2n}$ for the odd Sm isotopes near $N=90$, which indicates that
the first-order phase transition  also occurs in these odd Sm
nuclei. As shown in Fig.~\ref{F1}(b), the phase transition occurring
around $N=90$ corresponding to $N_p=4$ is explicitly manifested in
$S(2n)_{\mathrm{def}}$ for both the even and odd Sm isotopes. Thus
$N_p=4$ ($N=90$) may be considered as the critical point of the
phase transitions in the Sm nuclei. Particularly, the phase
transitional signal in the odd Sm nuclei seems to be greatly
enhanced by the odd particle effect. Specifically, the amplitude of
$S(2n)_{\mathrm{def}}$ in the odd Sm nuclei increases about $25\%$
near the critical point in comparison to that in the even Sm nuclei.

\begin{center}
\vskip.2cm\textbf{IV. Odd-even mass difference and pairing
excitation energy}
\end{center}\vskip.2cm

\begin{figure}
\begin{center}
\includegraphics[scale=0.22]{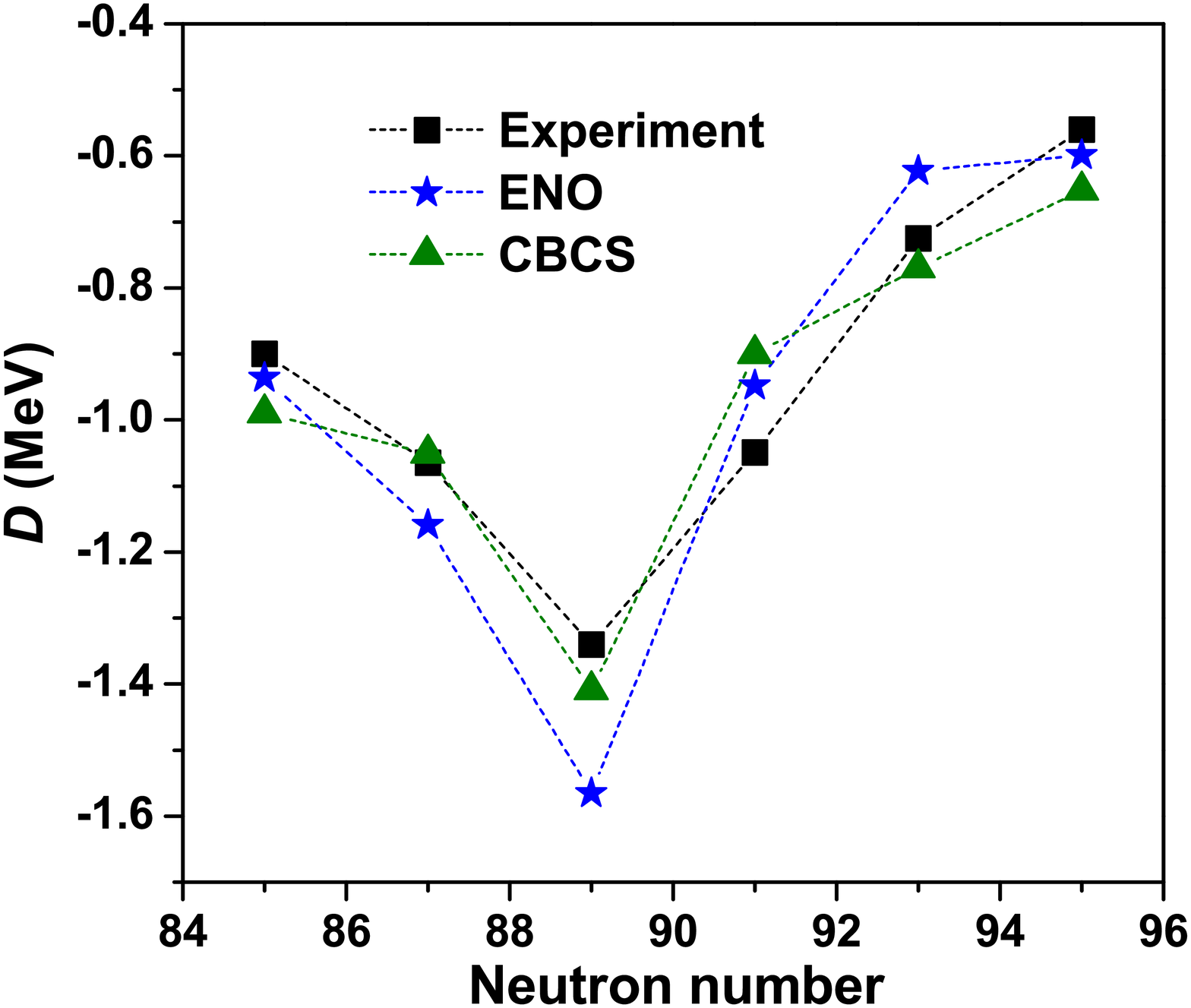}
\includegraphics[scale=0.22]{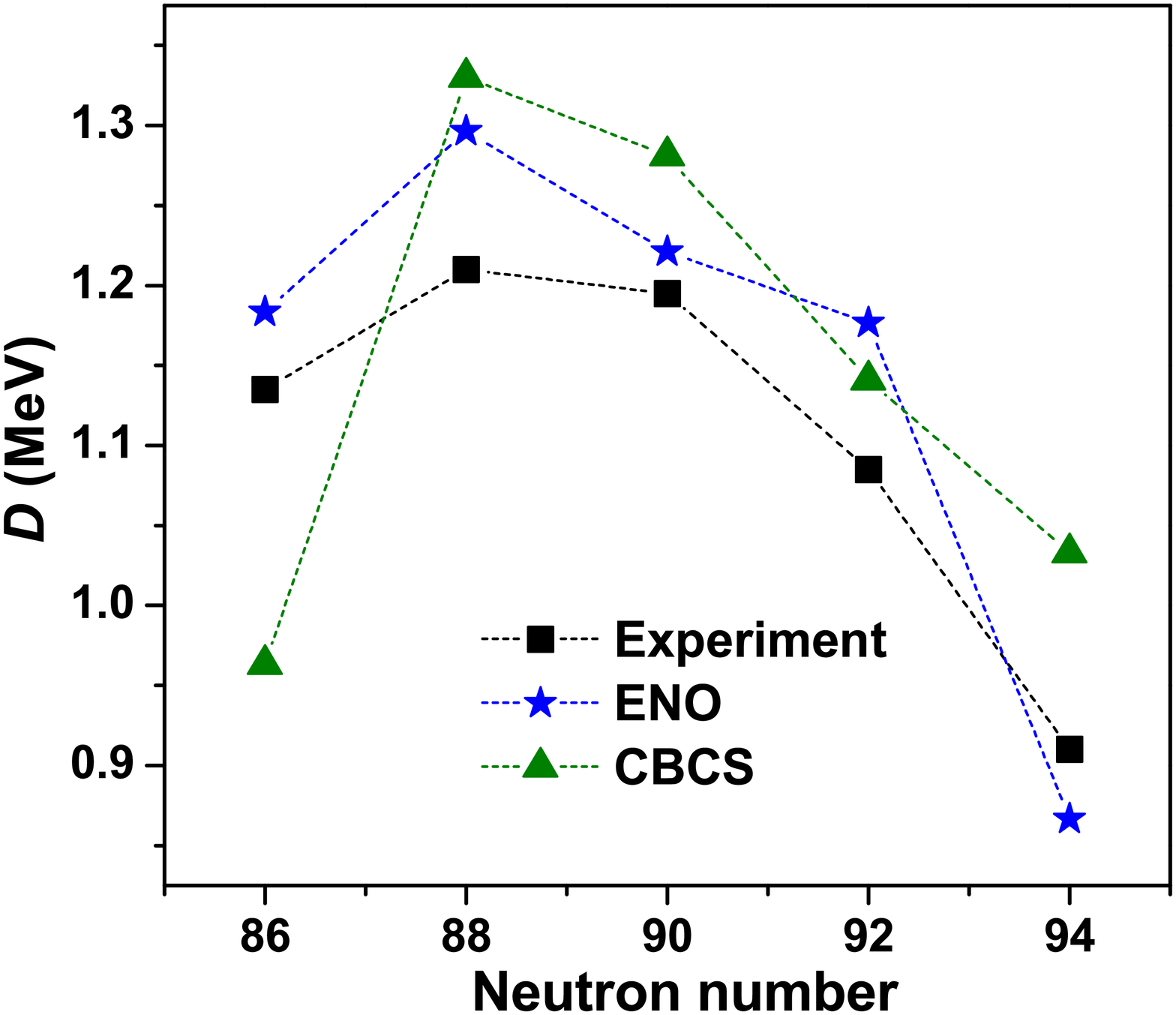}
\caption{(a) The odd-even mass difference $D$ for Sm nuclei with
neutron number $N=\mathrm{odd}$ fitted by the pairing model in the
two schemes; (b) the same as in (a) but for $N=\mathrm{even}$.
\label{F3}}
\end{center}
\end{figure}

Besides the two-neutron separation energy $S_{2n}$ and its deformed
part $S(2n)_{\mathrm{def}}$, there exist some other
binding-energy-related quantities that can serve as effective order
parameters to identify the shape phase transitions in odd
nuclei~\cite{Zhang2013}. A novel transitional signal in odd-even
nuclei related to the binding energy $B(Z,N)$ is given as the
odd-even mass difference~\cite{Zhang2013}, which is defined
as~\cite{Ring1980}
\begin{equation}
D=B(Z,N)-\frac{B(Z,N-1)+B(Z,N+1)}{2}\, .
\end{equation} Meanwhile, it is also well known that
the odd-even effects, such as those in the odd-even mass difference,
are the most important evidences of pairing in
nuclei~\cite{Ring1980}. Therefore, some microscopic factors relevant
to pairing in manifesting the phase transition can be extracted from
the odd-even effects. To do that, a shell model Hamiltonian
including the deformed mean field and pairing interaction will be
adopted to give a microscopic analysis of the phase transition
related to pairing in the Sm nuclei. Specifically, the Hamiltonian
is written as~\cite{Ring1980}
\begin{equation}\label{H}
\hat{H}=\sum_i \epsilon_i~(a_i^\dag a_i+a_{\bar{i}}^\dag
a_{\bar{i}})+\sum_{i,j} G_{ij}~ b_i^{\dag} b_j\, ,
\end{equation}
where $\epsilon_i$ represents the single-particle energy of the
$i$-th Nilsson level, and  $b_i^\dag=a_i^\dag a_{\bar{i}}^\dag$
($b_i=a_{\bar{i}}a_i$) is the pair creation (annihilation) operator
with $\bar{i}$ labeling the time-reversed state of that labeled by
$i$. To solve such a Hamiltonian, two approximation schemes are
considered for the pairing interaction in this work. One is the
nearest-orbit pairing-interaction model~\cite{Pan,Pan2002}, which is
a simplified version of the Gaussian-type pairing interactions
suitable for deformed nuclei~\cite{Molique1997} with the
orbit-dependent pairing strengthes written as
\begin{equation}\label{G}
G_{ij}=\alpha~e^{-\beta(\epsilon_i-\epsilon_j)^2}\, ,
\end{equation}
where $\alpha<0$ and $\beta>0$ are the adjustable parameters. It is
clear that the pairing strength $G_{ij}$ shown in (\ref{G}) is
orbit-dependent, and the nearer the two orbits the stronger the
pairing interaction between the two pairs. As an approximation to
the Gaussian-type interactions given in (\ref{G}), only the on-orbit
pairing interactions $G_{ii}$ and the nearest-orbit pairing
interactions $G_{ii+1}$ or $G_{ii-1}$ are considered in the
nearest-orbit pairing model, while $G_{ij}$ with $\vert i-j\vert\geq
2$ are neglected~\cite{Pan,Pan2002}. As a further approximations, we
set $G_{ii}=G_{ii\pm1}=G$. Thus, there is only one free-parameter
$G$ to be determined for each nucleus in the isotopes. Such a
pairing interaction form can be exactly solved for all the Sm nuclei
by directly diagonalizing the Hamiltonian in the valence nucleon
space. More details about the solutions of the nearest-orbit pairing
model can be found in Ref. \cite{Pan,Pan2002}. For convenience, we
denote the exactly solvable nearest-orbit pairing interaction as the
ENO scheme. Another approximation scheme is the constant pairing
interaction~\cite{Ring1980}, in which the paring strength is set as
$G_{ij}=G$ for all the single-particle orbits ${i}$. However, the
constant pairing interactions can be exactly solved only for a
nucleus with very few valence nucleons, due to the difficulty of
computation. To apply the constant pairing interactions form to the
whole chain of the Sm isotopes, the well known BCS
method~\cite{Ring1980} is used to solve the corresponding
Hamiltonian. We denote the constant pairing interaction solved by
the BCS theory as the CBCS scheme.
\begin{figure*}
\begin{center}
\includegraphics[scale=0.21]{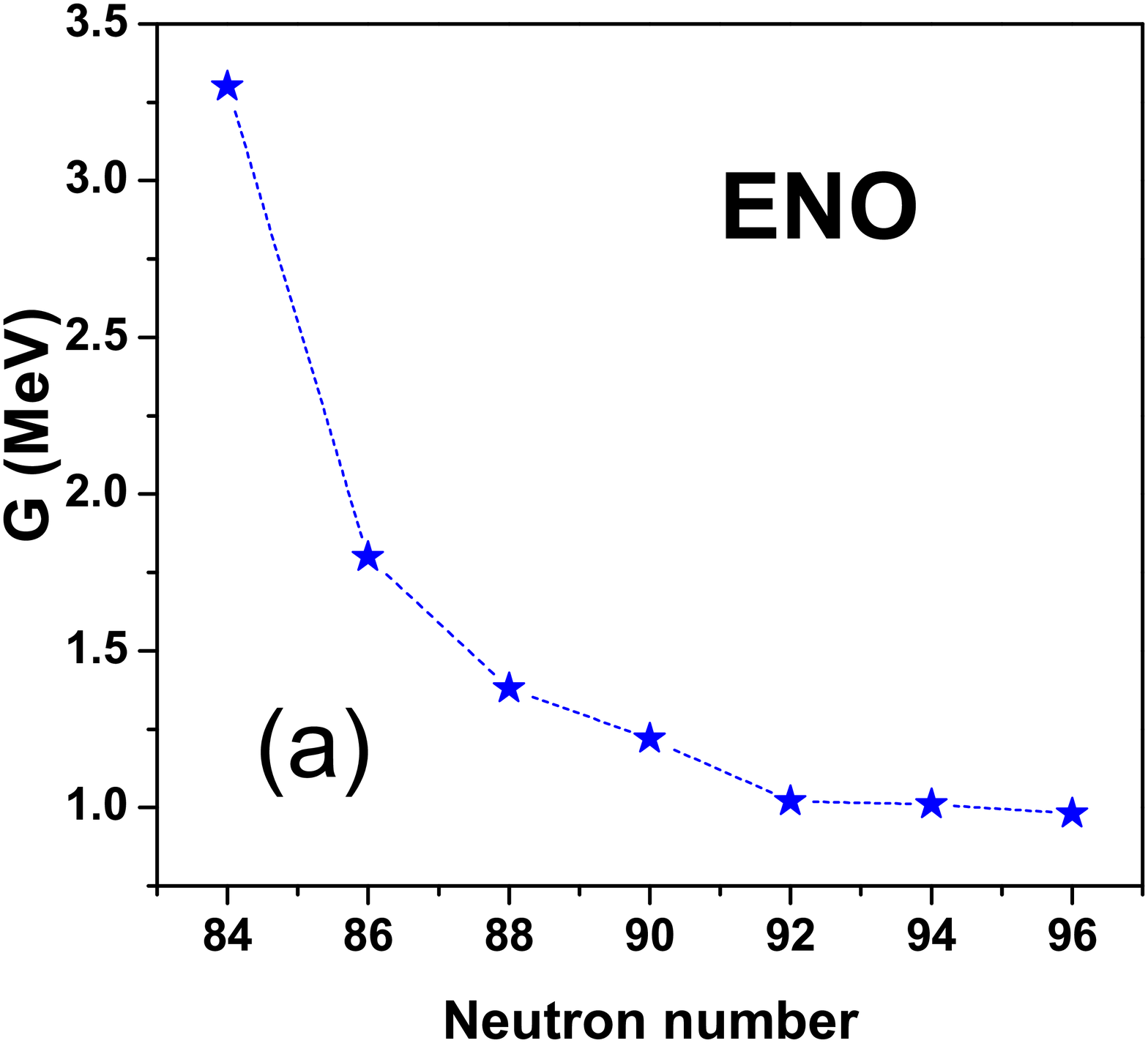}
\includegraphics[scale=0.21]{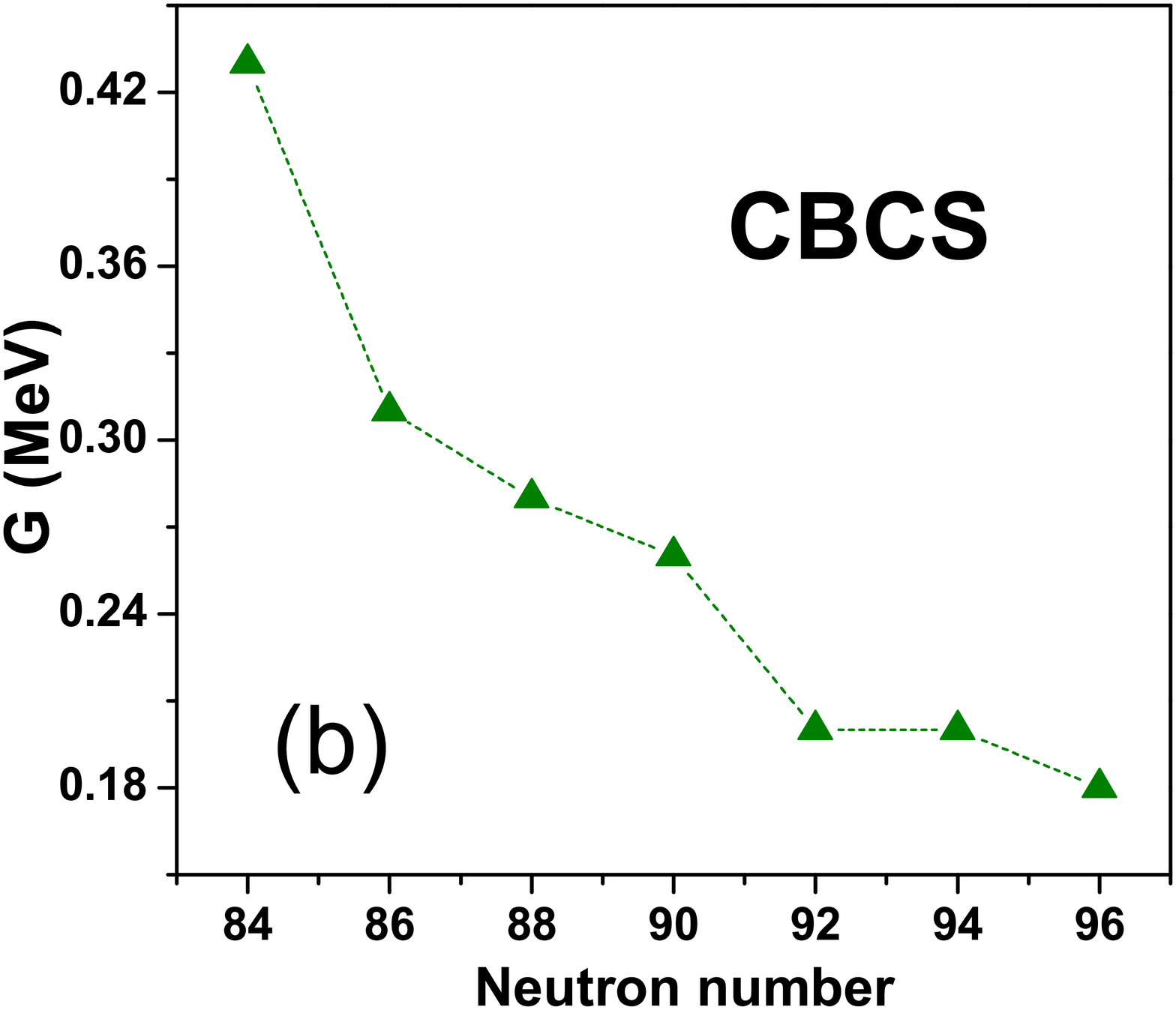}
\includegraphics[scale=0.21]{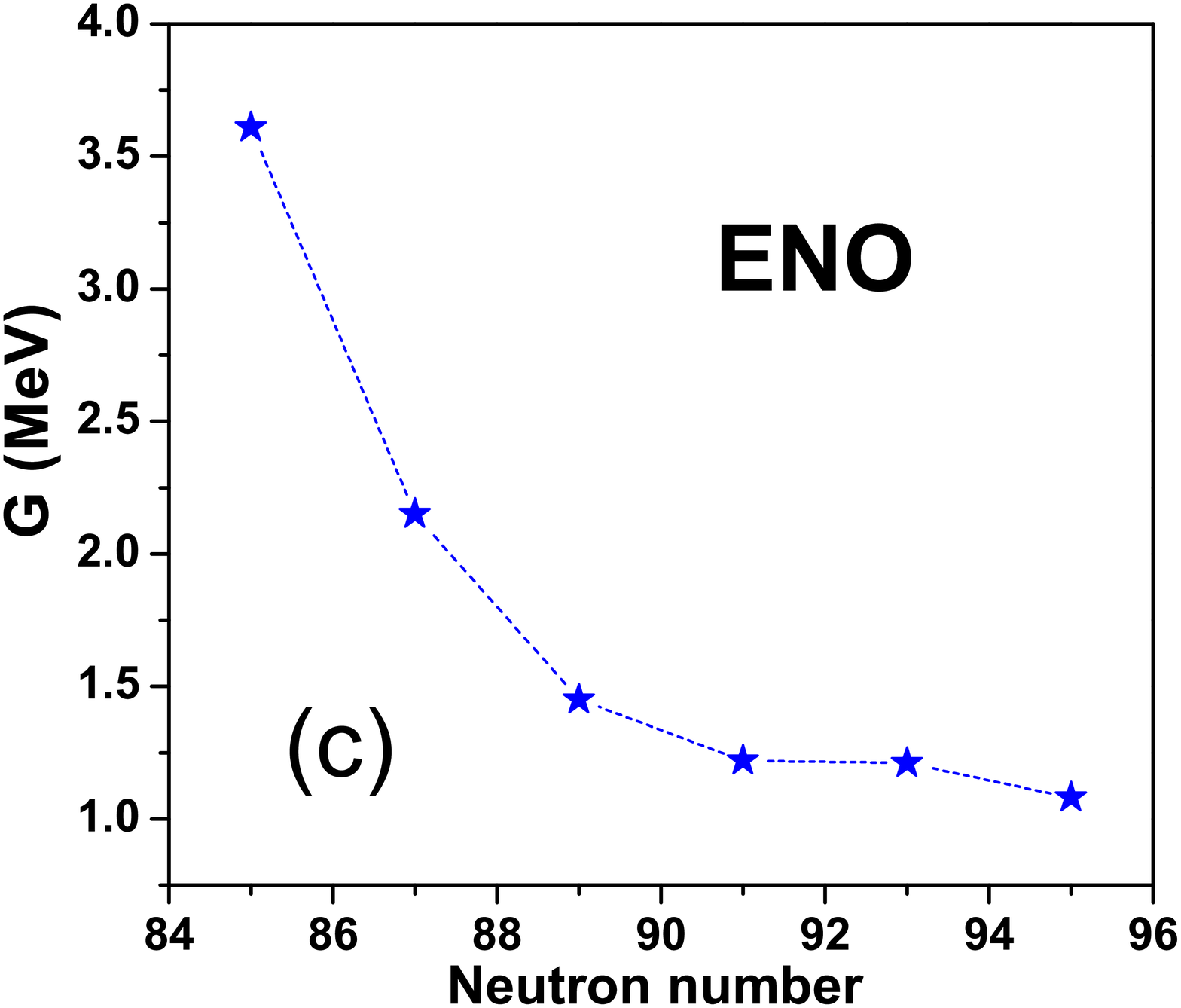}
\includegraphics[scale=0.21]{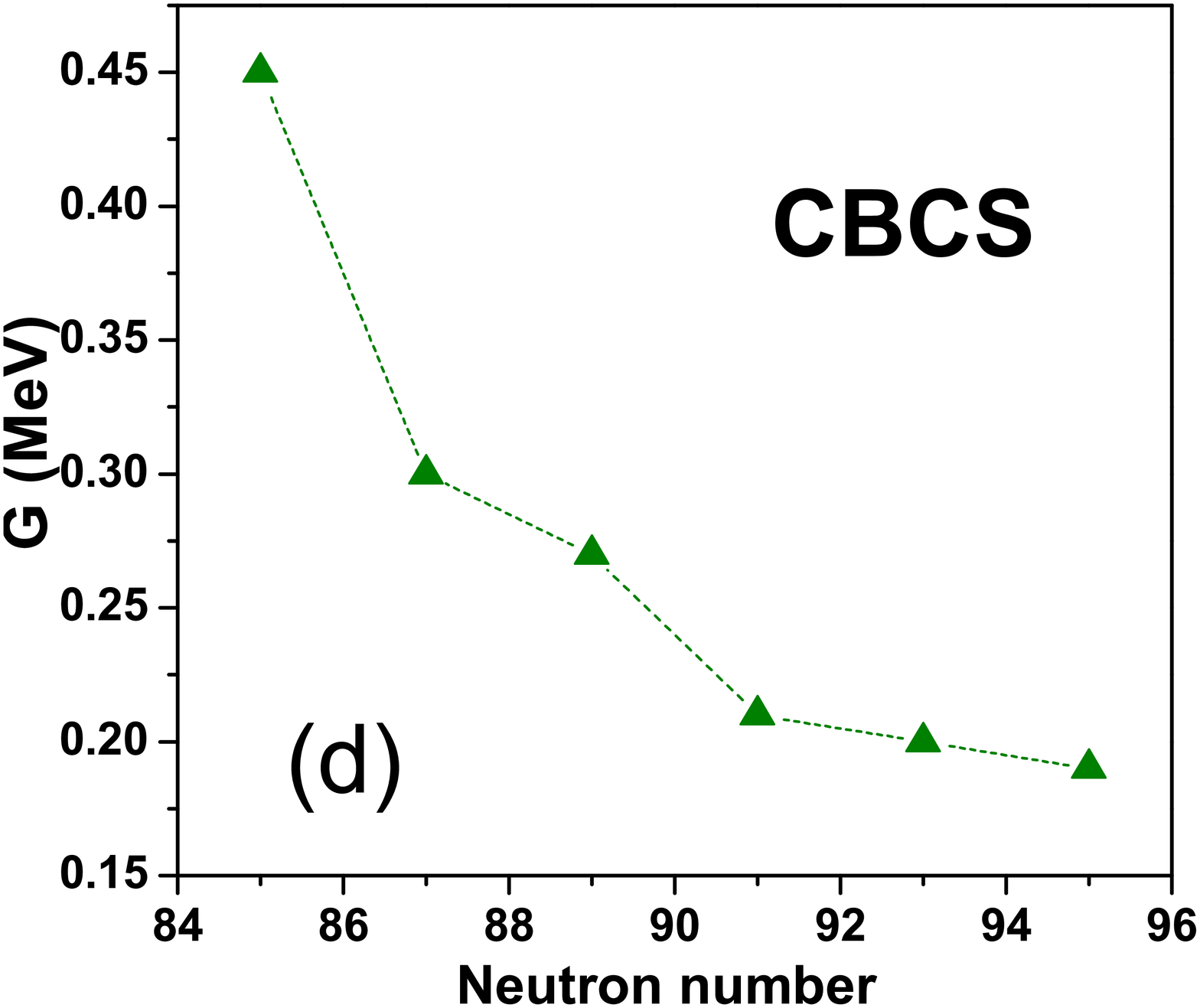}
\caption{The evolution of the pairing strength $G$ changing as a
function of neutron number. \label{F4}}
\end{center}
\end{figure*}

By using the ENO and CBCS schemes, the odd-even mass difference $D$
for the Sm nuclei has been fitted by the Nilsson mean-field plus the
pairing model and the resulting values together with the
experimental data are shown in Fig.~\ref{F3}. In theory, the
odd-even mass difference is given as
\begin{equation}
D=-E_g(Z,N)-\frac{-E_g(Z,N-1)-E_g(Z,N+1)}{2}\, ,
\end{equation}
where $E_g(Z,N)$ is the ground state energy solved from (\ref{H})
for a nucleus with the proton number $Z$ and the neutron number $N$.
 In our calculations, the
single-particle energies $\{\epsilon_i\}$ are calculated from the
Nilsson model with deformation parameters taken from \cite{mol},
which were determined systematically from the corresponding
experimental data~\cite{audi}. In addition, it is assumed that the
odd-even mass difference in the Sm isotopes only comes from the
neutron part since the proton number in a chain of isotopes is a
constant. As clearly seen from Fig.~\ref{F3}(a) and (b), the
experimental values of the odd-even mass difference in Sm can be
well reproduced in the pairing model for both schemes. Specifically,
the values of the odd-even mass differences $D$ are all negative for
the odd Sm nuclei but positive for the even Sm nuclei, which
indicates that even-even nuclei are more bounded than the odd-even
nuclei~\cite{Ring1980}. More important, the evident phase transition
signals in experiments shown by $D$, of which the values reach their
minimum or maximum around $N=90$, are nicely expressed by those
calculated from the pairing model. It is thus confirmed that the
pairing interaction is indeed a key factor in driving phase
transitions in nuclei.

Further, the fitted pairing strength $G$ in the two schemes is shown
in Fig.~\ref{F4}. As shown in Fig.~\ref{F4}(a) and (b), the
resulting pairing strength $G$ values in both schemes show a
monotonic decrease for the even Sm isotopes as the neutron number
$N$ increases, except that the variational behavior of $G$ in the
ENO scheme is a little smoother than that in the CBCS scheme. A
similar situation also appears in $G$ for the odd Sm nuclei as seen
in Fig.~\ref{F4}(c) and (d). It is thus confirmed that the phase
transition behavior related to pairing in an isotope may be driven
by the pairing interaction with a monotonic decrease in the pairing
strength as the neutron number increases. In addition, one may find
that the energy scale of $G$ in the ENO scheme is almost ten times
that in the CBCS scheme, as seen from Fig.~\ref{F4}. It is not
difficult to understand this by considering the fact that only the
on-orbit and nearest-orbits pairing interaction are taken into
account in the ENO scheme for a single-particle orbit $i$, in
contrast to the CBCS scheme. It seems that the the transitional
behavior of the odd-even mass differences $D$ can be well
illustrated in theory via the shell model Hamiltonian including only
the mean-field plus pairing interaction.

\begin{figure}
\begin{center}
\includegraphics[scale=0.2]{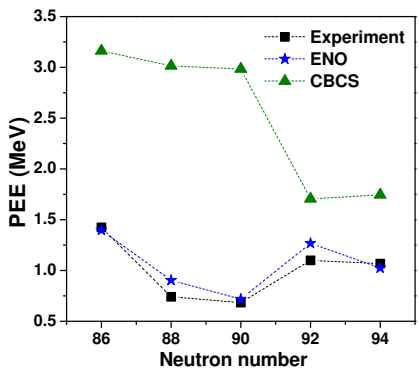}
\caption{The first pairing-excitation energy calculated by the two
schemes compared with that determined by
experiments~\cite{Internet}. \label{F5}}
\end{center}
\end{figure}

To further reveal the phase transitional behaviors closely related
to the pairing interaction, we also calculated the
pairing-excitation energy (PEE)~\cite{Pan2002}, which is also a
quantity sensitive to the pairing strength $G$. Since the angular
momentum projection along the third axis in the intrinsic frame is
considered to be a conserved quantity in the model, the
pairing-excitation states determined by the model are thus regarded
approximately as excited states with the same spin and parity as
those of the ground state of a nucleus. For example, the PEEs in the
model for even-even nuclei are considered as the energies of the
excited $0^+$ state, $E_{0_n^+}$. In Fig.~\ref{F5}, the results of
the first PEE calculated for the even Sm nuclei in the two schemes
and the corresponding experimental data are shown. Notably, only the
first neutron PEEs in the two schemes are taken to be compared with
the experimental data since the proton PEE may be much higher than
the corresponding neutron PEE in the present case. In the ENO
scheme, the first PEE may be briefly denoted as
$\mathrm{PEE}=E_1-E_g$, where $E_1$ and $E_g$ represent the energy
value of the first excited state and that of the ground state
respectively, since broken pairs  have not been taken into account
in this scheme. In contrast, the first PEE calculated from the CBCS
scheme can be explicitly given as
\begin{equation}\label{BCS}
\mathrm{PEE}=2\sqrt{(\varepsilon_0-\lambda)^2+\Delta^2}\, ,
\end{equation}
where $\lambda$ represents the Fermi energy, $\varepsilon_0$ denotes
the single-particle energy closest to $\lambda$, and $\Delta$ is the
so-called gap parameter~\cite{Ring1980}. All the parameters involved
in (\ref{BCS}) can be determined by the standard BCS
theory~\cite{Ring1980}. As clearly seen in Fig.~\ref{F5}, the first
PEE in experiments also provides an evident phase transitional
signal around $N=90$. More importantly, such a phase transitional
characteristic shown by the PEE can be well reproduced by the
results calculated from the pairing model in the ENO scheme. It is
thus further confirmed that the transitional characteristics in
connection with pairing in the Sm isotopes are indeed driven by the
pairing interaction with a monotonic decrease in the pairing
strength. However, it can be also found that the PEE obtained from
the constant pairing interactions in the CBCS scheme are much higher
than those found experimentally, and the global behavior of the PEE
in theory is also completely different from that in experiments. As
a consequence, the exact solutions are important to explore the
phase transition in the pairing model. It should be noted that the
PEEs in the odd Sm nuclei are not taken into account here because
some single-particle excitations with spin and parity the same as
those of the ground state are often involved in the low-lying
spectrum, which makes it difficult to pick out the PEE from the
spectrum of the odd Sm nuclei.

\vskip 1.3cm
\begin{center}
\vskip.2cm\textbf{VI. Summary}
\end{center}
\vskip.2cm

In conclusion, we have made a microscopic analysis of the shape
phase transition in the odd Sm nuclei from the point of view of the
effective order parameter. Through analyzing the two-neutron
separation energy, it is confirmed that the first order phase
transition also occurs in the odd Sm isotopes as it does in the even
Sm isotopes but with the signals of the phase transition in the odd
species greatly enhanced by the odd neutron effect. It is also shown
that the odd-even mass differences may reach their extreme value
around the critical point, thus serving as a valid effective order
parameter for the identification of the shape phase transition in
the odd Sm nuclei. Particularly, analysis based on the mean-field
plus pairing interaction Hamiltonian shows that the critical
phenomena relevant to pairing in the Sm nuclei can be driven by the
pairing interaction with a monotonic decrease in the pairing
strength $G$. In addition, the results also indicate that the
exactly solvable models are important to analyze transition
characteristics related to pairing in the excited states. Although
the discussion in this work provides a specific example of
microscopic analysis of the shape phase transition in odd-A nuclei,
investigations based on a more realistic shell model Hamiltonian
with different truncation
schemes~\cite{Sun2008,Zhao2000I,Zhao2000II} are still needed to
eventually confirm or disprove the theoretical predictions.

\bigskip

\begin{acknowledgments}
Support from the Natural Science Foundation of China (11375005,
11005056, 11175078 and 11405080).
\end{acknowledgments}

\end{document}